# Parfum: Detection and Automatic Repair of Dockerfile Smells

Thomas Durieux
TU Delft
The Netherlands
thomas@durieux.me

## ABSTRACT

Docker is a popular tool for developers and organizations to package, deploy, and run applications in a lightweight, portable container. One key component of Docker is the Dockerfile, a simple text file that specifies the steps needed to build a Docker image. While Dockerfiles are easy to create and use, creating an optimal image is complex in particular since it is easy to not follow the best practices, when it happens we call it Docker smell. To improve the quality of Dockerfiles, previous works have focused on detecting Docker smells, but they do not offer suggestions or repair the smells. In this paper, we propose, Parfum, a tool that detects and automatically repairs Docker smells while producing minimal patches. Parfum is based on a new Dockerfile AST parser called Dinghy. We evaluate the effectiveness of Parfum by analyzing and repairing a large set of Dockerfiles and comparing it against existing tools. We also measure the impact of the repair on the Docker image in terms of build failure and image size. Finally, we opened 35 pull requests to collect developers' feedback and ensure that the repairs and the smells are meaningful. Our results show that Parfum is able to repair 806 245 Docker smells and have a significant impact on the Docker image size, and finally, developers are welcoming the patches generated by Parfum while merging 20 pull requests.

## 1 INTRODUCTION

Docker is a popular tool for developers and organizations to package, deploy, and run applications in a lightweight, portable container. One key component of Docker is the Dockerfile, a simple text file that specifies the steps needed to build a Docker image. While Dockerfiles are easy to create, creating the optimal Docker image is a complicated task especially when the best practices are not the same outside the Docker environment. When a best practice is not followed, we call it a Docker smell. Docker smells are particularly frequent inside Dockerfiles because most developers that wrote them are not experts [13]. Moreover, the best practices that are used in interactive Bash differ from the best practices that are used in Bash inside a Dockerfiles which leads to non-optimal Docker images.

Previous works from academics and the industry have focused on detecting Docker smells. For example, Binnacle [13], hadolint [11], dockerfilelint [3], docker-bench-security [2] or dockle [4] are linters designed to identify a large variety of Dokcer smells. Unfortunately, those tools are not well known or are not maintained in the case of dockerfilelint and dockle. Additionally, none of those tools offer suggestions or repair the smells. Which makes it harder for the developers to fix or understand how those smells could be removed. Having a repair technique not only automatizes the repair but also provides examples of how the smell can be repaired. This increases the developer's knowledge and reduces the repair overhead.

In this paper, we target the gap between the detection and the repair of Docker smells by proposing Parfum. Parfum detects and automatically repairs Docker smells by producing minimal patches. Parfum is based on a new AST parsing tool called Dinghy that has been designed to simplify Dockerfile refactoring and allows transforming Dockerfiles without affecting the formatting of the file. This feature is particularly important if we want that the tool is used by the developers. The detection of smells follows a similar methodology then Binnacle [13] and makes it more precise and extends it to support a wider range of smells and finally makes the tool usable by developers. For the repair, Parfum uses a template-based approach to repair the smells. The advantage of the template-based approach is to generate quickly precise patches that look identical to developer's patches.

We evaluate the effectiveness of Parfum by analyzing and repairing 369 073 Dockerfiles. We compare Parfum against Binnacle [13] using the 178 452 Dockerfiles from Binnacle evaluation. In this comparison, we highlight the limitations of Binnacle and how Parfum is targeting those limitations. We follow this comparison by evaluating the ability of Parfum to repair the 806 245 Docker smells that Parfum detected. Additionally, we selected 13 440 Dockerfiles that contain at least one smell that we built with and without the repairs to check if Parfum introduces regressions. We also measure the impact of the Docker smells on the resulting Docker image sizes. Finally, we opened 35 pull requests with the output of Parfum to collect developers' feedback and ensure that the repairs and the smells are meaningful. During this evaluation, we show that Parfum is able to repair Docker smells that are frequently present in Dockerfiles and that have a significant impact on the size on the image sizes that we measure to be on average 46,38 MB per Dockerfile. Importantly, those images are frequently downloaded and those smells introduce a traffic of 39,93 TB per week. 20 (100,0 %) pull requests have been merged saving a total of 2,68 GB or 1,37 TB per week which indicates that developers are interested in Docker smells repair.

The potential impact of this contribution is significant. By automatically detecting and repairing Docker smells of good practices in Dockerfiles, we can help improve the quality and efficiency of Docker images. Additionally, by educating developers about best practices for Dockerfile development, we can help improve the overall quality and maintainability of Dockerfiles, which can lead to more robust and efficient containerized applications.

In summary, the contributions of this paper include:

- Dinghy, a library to analyze, modify, and print Dockerfile AST,
- Parfum, a tool that detects and repairs 32 Docker smells,
- A new dataset of 193 948 Dockerfiles extracted from GitHub,
- A ground truth dataset of 384 Dockerfiles manually annotated with the 23 smells,
- A comprehensive empirical study of the effectiveness of Parfum to repair Docker smells,
- 35 pull requests on popular GitHub repositories that repair 62 Dockerfile smells.

## 2 BACKGROUND

In this section, we provide background information on several key terms related to this contribution.

**Docker** is a tool designed to create, deploy, and run containerized applications.[1] Containers allow developers to package an application and its dependencies into a single, isolated package that can be easily deployed. This makes it easier to build, test, and deploy applications in a consistent and predictable way, regardless of the environment in which they are run. Docker is, for example, used to support reproducible research [1].

**Docker Image** is an executable package for Docker that includes everything needed to run a piece of software, including the code, a runtime, libraries, environment variables, and config files. Docker images are built using instructions contained in a Dockerfile.

**Dockerfile** is a text file that contains instructions for building a Docker image. The instruction define the base image to use (`FROM <image>`), the files to include (`COPY <source> <dest>`), the ports to open (`PORT <port>`), the entry point (`ENTRYPOINT <script>`), and the scripts to execute (`RUN <script>`). The scripts are interpreted by the shell of the image which is generally bash or PowerShell (for Windows Docker image).

**Docker Smell** refers to a potential issue or problem with a Dockerfile or Docker image [24]. This issue is generally detected when the Dockerfile or image are violating some rules. Common Docker smells include bloated images, misconfiguration, and misuse of commands. Identifying and addressing these smells can help improve the efficiency, security, and maintainability of a Docker-based project. In this paper, we focus on Dockerfiles and therefore we limit the scope of the smells to Dockerfiles.

## 3 CONTRIBUTION

In this section, we present, Parfum, a tool to automatically repair Docker smell. Parfum is based on Dinghy, a new library that is designed to parse, navigate, modify and reprint Dockerfiles. Dinghy has been created with the mind of simplifying Dockerfile refactoring. Parfum, on the other hand, is designed to be easy to use and easily extendable to support additional Docker smells. We will first present Dinghy in Section 3.1 and Parfum in Section 3.2.

### 3.1 Dinghy

Dinghy is a library to parse, query, manipulate and reprint Dockerfiles written in 3,68 k lines of TypeScript. It has been designed to make it easy to refactor Dockerfiles by lowering the entry costs of analyzing and transforming Dockerfiles. Dinghy is written in TypeScript to be compatible with Visual Studio Code, one of the most popular IDE[2] and with the browser to increase its adoption and all tools that will depend on it, such as Parfum and its web demo https://durieux.me/docker-parfum.

Dinghy relies on two existing libraries to parse Docker instructions and Bash script: dockerfile-ast and mvdan-sh. The two ASTs are then combined into a unified and consistent AST that can be used to manipulate Docker and Bash nodes transparently. In total, we define 141 node types, 100 bash nodes, 41 Docker nodes. Dinghy and its documentation are available on GitHub [5].

```
parse.ts

const ast = await dindhy.dockerParser.parseDocker(`FROM ubuntu:latest

RUN apt-get update && \
  apt-get install -y \
    curl;
ENTRYPOINT ["curl"]`);
ast.toString();
// FROM ubuntu:latest
// RUN apt-get update && apt-get install -y curl;
// ENTRYPOINT ["curl"]

ast.toString(true);
// FROM ubuntu:latest

// RUN apt-get update && \
//   apt-get install -y \
//     curl;
// ENTRYPOINT ["curl"]
```

**Figure 1: Dinghy pretty print feature.**

```
parse.ts

const ast = await dindhy.dockerParser.parseDocker(`FROM ubuntu:latest
  RUN apt-get update && \
    apt-get install -y curl;
  ENTRYPOINT ["curl"]`);
  ast.getElements(BashCommandArgs);
  ast
    .getElement(MaybeSemanticCommand)
    ?.getParent(BashScript)
    ?.getChild(MaybeSemanticCommand)
    ?.getChildren(BashCommandArgs);
  // iterate over the children
  ast.iterate(
    (node) => {
      // do something with the node
    },
    (node) => node instanceof DockerRun
  );
  // iterate over all nodes that have a value
  ast.traverse((node) => {
    if (node instanceof DockerOpsValueNode) {
      node.value;
    }
  });
  // search for command with at least 2 arguments
  ast.find(Q(MaybeSemanticCommand, Q(BashCommandArgs, BashCommandArgs)));
```

**Figure 2: Query features of Dinghy.**

Dinghy has a strong focus on refactoring which means that the AST should be easily manipulated but also that the code transformation does not break the code formatting of the Dockerfile. Dinghy is able to reprint the Dockerfile AST with minimal changes compared to the original formatting. We call this feature pretty print. This feature is illustrated in Figure 1. We handle this complex problem by only reprinting the nodes that were modified and otherwise reusing the content that was written by the developers. To be able to do so, we need to track accurately the modifications that have been done on the AST and have accurate positions of each node to be able to find the textual representation of the node in the original Dockerfile. Unfortunately, the positions were not always accurate which reflected poor performance of the pretty print before fixing the positions. We also had to unify the position between the Dockerfile AST and the Shell AST since they were parsed separately. Additionally, all white spaces (indentations, new lines, and spaces) of the Dockerfiles are reprinted following common practices which produce reliable results according to our tests.

[1] Docker: https://www.docker.com
[2] https://pypl.github.io/IDE.html visited on February 9, 2023

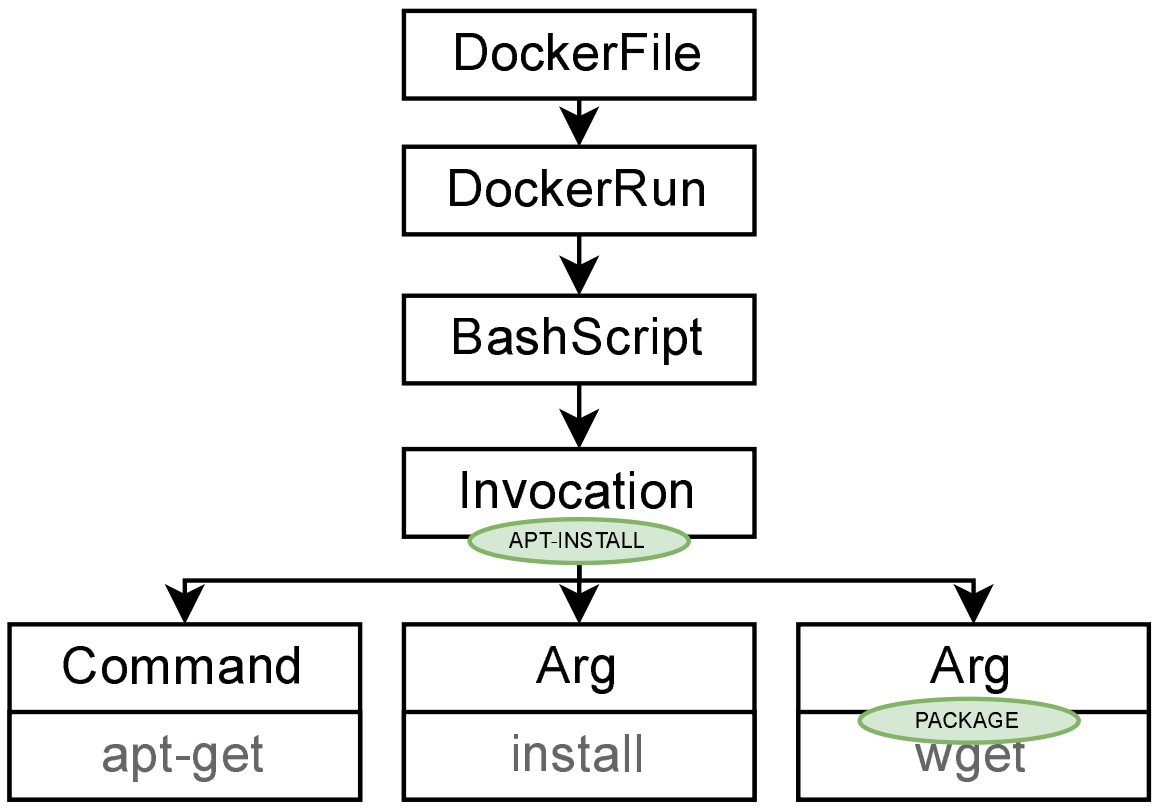

**Figure 3: Example of Dinghy AST annotated with structural information of the apt command. The green ovals are the annotations in the AST.**

Being able to navigate the AST is one of the most important features of an AST library. Dinghy supports the traditional navigation features such as getting the parent and the children but it also supports, as presented in Figure 2, a query system. The query allows finding nodes that match a specific pattern. The query system uses the node types but also annotations and node values. In Dinghy, it is indeed possible to annotate each node which is particularly useful when a multiple steps analysis is required. The results from the previous analysis can be used in the following ones. This allows for doing complex searches with relatively simple queries. This will be extensively used by Parfum.

## 3.2 Parfum

Parfum is a tool for detecting and repairing Docker smells in Dockerfiles. As previously mentioned, Parfum is based on Dinghy, and offers rules to identify Docker smell and repair them inside the Dockerfile AST.

The core methodology for detecting the Docker smell is inspired by Binnacle, a contribution by Hankel et al. [13]. The main idea is to enrich the Docker AST to include the structural information from command lines. For example, if we consider `RUN apt-get install wget` and its AST representation in Figure 3, the enriched AST contains information that specifies that `apt-get` is used to `install` packages and that the installed package is `wget`. In other words, the node that contains `apt-get install wget` is annotated with `APT-INSTALL`, the node `wget` with `PACKAGE` see the green ovals in Figure 3. As presented in Section 3.1, those annotations are part of the query system of Dinghy and therefore can be used inside the queries. For example, if we want to verify that all `apt-get` commands have the `-y` flag, we can look for all `apt-get` commands `ast.find(Q("APT-GET-INSTALL"))` and then verify that the command does not have the `-y` flag: `!apt.find(Q("APT-GET-INSTALL-Y-FLAG"))`.

Parfum currently supports 70 command lines which represent 89,05 % of all the commands present in the Dockerfiles of our dataset. We extended the list of supported commands compared to Binnacle's contribution to cover additional use cases and smells. Additionally, Parfum is also able to enrich commands that are embedded inside other commands. For example, the command `sudo apt update` contains a main command (`sudo`) and an embedded command `apt update`. Without supporting the embedded commands, Parfum would not be able to enrich an important part of the AST which frequently contains smells as we will see during the evaluation of Parfum (see Section 4.3).

We implemented in Parfum 32 Docker smell detections, 23 were implemented in Binnacle [13], 6 rules from hadolint [11] and 3 are original from this work. The complete list of supported smells is available in the repository of Parfum [7]. The list of supported rules with be extended in the future, the current support demonstrates the feasibility of the repair and evaluates the impact and relevance for the developers. Listing 1 presents an example of how the smell detections are implemented inside Parfum. A query defines which AST nodes to be present inside the AST to trigger the rule. The query can be refined with post-condition to specify if additional AST nodes need to be present before, after, or inside the node. In the example, we define a query that is looking for the command `npm cache clean` using the query `Q("NPM-CACHE-CLEAN")`. The post-condition verifies that the flag `-f` is not present inside the node. If those two queries have a match, Parfum has detected the smell and it is reported to the developer.

Once a smell is detected, Parfum repairs it by modifying the AST. Listing 2 contains an example on how the Parfum fixes a smells. In this example, Parfum adds the flag `--force` as an argument of the command `npm cache clean`. Once the AST is modified, Parfum reprints the AST to provide a repaired version of the Dockerfile. The reprint uses the pretty print feature of Dinghy and therefore only reprints the nodes that have been modified. An example of such transformation is visible in Listing 3.

In summary, the overall process followed by Parfum to detect and repair Docker smell is divided into five steps as follows:

(1) First, Parfum uses Dinghy to parse the Dockerfile, creating a unified AST representation of the Dockerfile.
(2) Parfum enhances the AST to include information about 70 command lines.
(3) Once the AST is enriched, Parfum uses a set of queries (illustrated in Listing 1) to search for smells in the AST.
(4) If a smell is detected, Parfum repairs it by modifying the AST (as shown in Listing 2).
(5) Parfum reprints the modified AST, creating a new Dockerfile with the smells repaired (as shown in Listing 3).

Finally, Parfum is available on GitHub [7] and it also has been ported to a browser version which is available at https://durieux.me/docker-parfum. We plan to continue to develop Parfum to support additional Docker smells. We also plan to estimate the size impact of the repairs statically in order to provide a quick estimation to the developers. We also plan to extend it for doing advance refactoring such as transforming a single-stage to a multi-stage Dockerfile.

# 4 EVALUATION

In this section, we present the evaluation of Parfum, where we aim to assess the effectiveness of Parfum in detecting and repairing Docker smells. We also study the impact of the smells on Docker image size and how developers react to the Docker smell repairs.

```
{
 // search for `npm cache clean`
 query: Q("NPM-CACHE-CLEAN"),
 consequent: {
  // verify if `-f` flag is inside the node
  inNode: Q("NPM-F-FORCE"),
}}
```

**Listing 1: Parfum query to detect *npmCacheCleanUseForce* smell.**

```
function repair(node) {
  // insert the --force arguement
  node.addChild(new BashCommandArgs().addChild(
    new BashLiteral("--force")
  ));
}
```

**Listing 2: Parfum code to repair *npmCacheCleanUseForce* smell.**

```
@@ -21,1 +21,1 @@
-RUN npm cache clean
+RUN npm cache clean --force
```

**Listing 3: Parfum diff of the repaired *npmCacheCleanUseForce* smell.**

## 4.1 Methodology

To evaluate the effectiveness of Parfum, we design and conduct an empirical evaluation to answer the following research questions:

RQ1 **What is the effectiveness of Parfum in detecting Docker smells?** To answer this question, we analyze with Parfum two datasets containing a total of 372 400 Dockerfiles, those two datasets are presented in Section 4.2. One of the datasets is presented by Hankel et al. [13] when introducing Binnacle which we use to compare the effectiveness with Parfum. We follow the comparison with precision and recall analysis on a new ground truth dataset. We end the research question by highlighting the differences and improvements that bring Parfum compared to Binnacle.

RQ2 **What is the effectiveness of Parfum in repairing Docker Smells?** In this second question, we evaluate the ability of Parfum to repair Docker smells. To do so, we conduct a quantitative analysis of the repair of Docker smells where we repair using Parfum the 210 901 Dockerfiles that contain at least one smell. Then we analyze again the repaired Dockerfiles with Parfum to verify that the smell disappeared. Furthermore, we selected 13 440 smelly Dockerfiles (see Section 4.2) and built them to ensure that the repairs do not break the Docker build.

RQ3 **What is the impact of the Docker smell repairs on the Docker image size?** In this third research question, we study the impact of the Docker smells on the size of the Docker images. The goal of this research question is to measure the impact of the Docker smells. To answer this question, we measure the size before and after the repair for the 6 138 Dockerfiles that are buildable after the automatic repair of Parfum. We also measure the theoretical gain in terms of bandwidth on Dockerhub if all the repairs are applied.

RQ4 **Do developers accept the Docker Smell repairs?** In the final research question, we aim to evaluate the interest that the developers have in the repair of Docker smells. To do so, we opened 35 pull requests to propose the repairs generated by Parfum to the developers. The code modification is completely automatized and was not modified by a human however the body and the title of the pull request were adapted to fit the contribution guidelines of the repositories. We selected the projects based on the activity, contribution requirements of the repository, the number of pulls on Dockerhub, the number of repairs, and the size reduction.

By addressing these research questions, we aim to provide a comprehensive evaluation of the effectiveness of Parfum in detecting and repairing Docker Smells and understanding their impact on the development process and the developers' interest in Dockerfile smell repairs.

## 4.2 Dataset

In this section, we present the datasets that we use in the evaluation of Parfum. The first dataset used in this paper is Binnacle [13], which contains 178 452 unique Dockerfiles extracted from GitHub repositories in 2020. This dataset will be used in the first research question to compare the Docker smell detection effectiveness of Binnacle with our proposed tool, Parfum.

We also created a new dataset of 193 948 Dockerfiles that were extracted from GitHub repositories in 2022 (two years after the Binnacle dataset). In this dataset, we have included important information such as the origin repository, commit SHA, and the path of the Dockerfile. The main difference between the new dataset and Binnacle Dataset is when the two datasets have been created and that the new dataset contains the origin of the Dockerfiles. It is therefore possible to link a Dockerfile to a repository that allows future studies, such as RQ3 (see Section 4.5) and RQ4 (see Section 4.6). Table 1 summarizes the main characteristics of the two datasets and presents some of their differences.

We will now present the methodology used to create this new dataset. The first step in creating this dataset was to identify a set of potential repositories. We decided to select repositories that were 1) not forks, 2) had at least 10 stars, and 3) had at least 50 commits. We choose those criteria in order to obtain Dockerfile from repositories that have a minimum of activity and that are more likely to have been maintained. We ended up with a list of 500 108 potential repositories.[3]

The next step was to download the file list from the default branch of the latest commit for each repository. We were able to download the file list for 500 022 repositories.

The final step was to identify and download the Dockerfiles stored in these repositories. We iterated over the list of files and considered any files that contained the string "Dockerfile" (case sensitive) as potential Dockerfiles. This resulted in a collection of 193 948 Dockerfiles that constitute the new Dataset which is available on our online artifact [6].

## 4.3 RQ1: Docker smell detection

In this first research question, we study the effectiveness of Parfum to identify Docker smells in a large set of Dockerfiles. We will compare the smell detection of Parfum with Binnacle by Hankel et al. [13] on their dataset. We will also perform the analysis on our new dataset and compare the results between the two datasets.

[3]Downloaded on July 12, 2022 from https://seart-ghs.si.usi.ch/

**Table 1: Characteristics of Binnacle [13] and Parfum datasets.**

| Metric | Binnacle [13] | Parfum |
|---|---|---|
| Creation date | 2020 | July 2022 |
| # Dockerfile | 178 452 | 193 948 |
| # Smelly Dockerfile | 100 876 | 110 487 |
| Total # Instruction | 2 223 139 | 3 637 952 |
| Avg. # Instruction Dockerfile | 12,45 | 18,04 |
| Med. # Instruction Dockerfile | 9 | 12 |

Table 2 presents the comparison of the detection between Binnacle and Parfum. The first column contains the name of the Docker smell as defined by Hankel et al [13]. The second column contains the number of Docker smell occurrences detected by Binnacle. The third column contains the number of Docker smell occurrences detected by Parfum. The fourth column contains the difference between the two tools. The fifth column contains the number of Dockerfiles with smell detected by Binnacle. The sixth column contains the number of Dockerfiles with smell detected by Parfum. The last column contains the difference between the two tools. The last row contains the total of detected Docker smells and to sum of the difference between the two tools, i.e., the sum of the absolute values of the "Diff" columns.

A total of 68 015 differences has been identified between the two analyses which is a significant difference considering that the two tools follow the same detection methodology. In summary, Parfum and Binnacle identify smells in a third of the files. Binnacle identifies smells in more files but fewer smells in total.

To measure the precision of Parfum and Binnacle, we create a ground truth dataset of 384 Dockerfiles that are randomly selected from the Binnacle Dataset. We chose a sample of 384 Dockerfiles to have a confidence level of 95% that the real value is within ±5% of the full dataset. Parfum has an average a precision of 0,96, recall 0,95, F1 of 0,94 and a median precision of 1, recall 1, F1 of 1. On the other hand, Binnacle has an average precision of 0,93, recall 0,918, F1 of 0,91, and a median precision of 1, recall 1, F1 of 1. Based on those results, we concluded that Parfum is more precise than Binnacle to detect this set of Docker smells. Several bugs that have been identified while comparing the results with the manual analysis have been repaired and therefore even further improve the presented results. We do not include the updated results to keep a fair comparison.

We identify several reasons that explain this difference between the two tools. The first reason is that Parfum detects is able to detect smells inside sub-commands such as `sudo`. Indeed, a Dockerfile that contains `sudo apt-get install docker` is considered as smelly by Parfum but not by Binnacle (one of the smells is a missing `-y` flag). This reason explains the majority of the difference in the *aptGet** rules.

The second main reason is the lack of precision of the Binnacle queries. For example, in the case of *wgetUseHttpsUrl* and *curlUseHttpsUrl*, Binnacle only checks the absence of a string containing `https` in the child nodes of `wget` or `curl`. However, this trigger false positive when a different protocol is used or when a variable is used such as in those cases: `wget ftp://google.com/all_data.zip`, `wget $HTTPS_URL`. On the other hand, Parfum would only trigger that rules when the HTTP protocol is used.

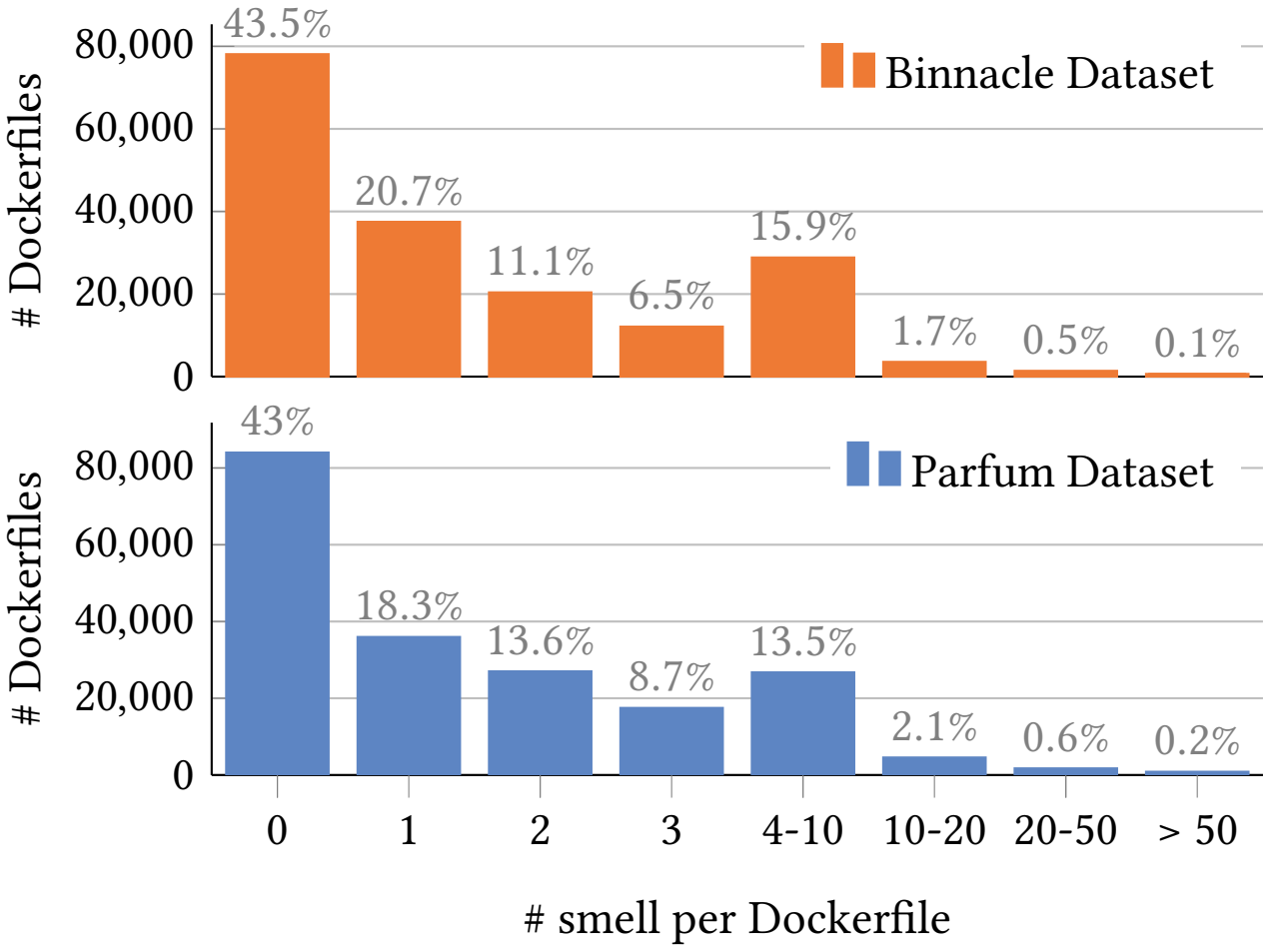

**Figure 4: The distribution of the number of smells detected by Parfum per Dockerfile on the two datasets.**

The third reason is that Parfum takes into consideration conditions, i.e., `if`, which impacts the results of the analysis in a few cases. Fortunately, not many Dockerfiles contain complex shell scripts and therefore not many of them are impacted by this issue. Unfortunately, conditions have a bigger impact on the repair when deciding where the repair should be placed. The final reason is due to bugs in Binnacle implementation which leads to invalid AST annotations.

Finally, We compare the results on Binnacle Dataset with the new dataset. Figure 4 presents the distributions of the number of smells per Dockerfile in Binnacle and the new dataset respectively. The two figures are relatively similar however Parfum Dataset seems to have slightly more Dockerfiles with two and three smells while the Binnacle dataset has more Dockerfiles with one smell. While the number of Dockerfiles without smell slightly increased as observed by previous studies [8, 18]. The increase in the number of smells per Dockerfile is probably due to the fact that the number of instructions per Dockerfile has increased by 1/3 between the two years from an average of 12,45 to 18,4 (median 9 and 12) respectively.

**Answer to RQ1**. We show that Parfum is able to detect 806 245 Docker smells inside the two considered datasets and took on average 380ms per analysis. We show that Parfum has an average precision of 0,96, recall of 0,95, and an F1 score of 0,94 on our ground truth dataset. Finally, we discussed several reasons why Parfum outperforms Binnacle in the detection of Docker smells.

## 4.4 RQ2: Effectiveness of automatic repairing of Docker smells

In this second research question, we are looking at the effectiveness of Parfum to repair Docker smells. To answer this research question, we use Parfum to automatically repair the smells in the 210 901 Dockerfiles that contain at least one smell. We then verified that the smells were fixed by analyzing the repaired Dockefiles. Then, for a selection of Dockerfiles, we build the Docker image to ensure that the repair did not break the build.

**Table 2: The number of smells and smelly Dockerfiles detected by Binnacle [13] and Parfum on the Binnacle dataset.**

| Docker Smell | # Docker Smell | | | # Dockerfile with Smell | | |
|---|---|---|---|---|---|---|
| | Binnacle [13] | Parfum | Diff | Binnacle [13] | Parfum | Diff |
| aptGetInstallUseNoRec | 61 836 (18,2 %) | 78 810 (20,6 %) | 16 974 | 36 770 (36,6 %) | 40 825 (40,5 %) | 4 055 |
| aptGetInstallThenRmAptLists | 57 079 (16,8 %) | 73 921 (19,3 %) | 16 842 | 32 349 (32,2 %) | 36 250 (35,9 %) | 3 901 |
| curlUseFlagF | 50 613 (14,9 %) | 48 423 (12,6 %) | −2 190 | 36 968 (36,8 %) | 36 819 (36,5 %) | −149 |
| pipUseNoCacheDir | 29 735 (8,8 %) | 30 133 (7,9 %) | 398 | 16 568 (16,5 %) | 16 793 (16,6 %) | 225 |
| gpgUseBatchFlag | 29 127 (8,6 %) | 29 110 (7,6 %) | −17 | 10 070 (10,0 %) | 10 069 (10,0 %) | −1 |
| aptGetUpdatePrecedesInstall | 27 017 (8,0 %) | 37 942 (9,9 %) | 10 925 | 13 151 (13,1 %) | 16 517 (16,4 %) | 3 366 |
| npmCacheCleanAfterInstall | 14 149 (4,2 %) | 14 189 (3,7 %) | 40 | 10 467 (10,4 %) | 10 486 (10,4 %) | 19 |
| yumInstallRmVarCacheYum | 12 247 (3,6 %) | 12 277 (3,2 %) | 30 | 6 102 (6,1 %) | 6 121 (6,1 %) | 19 |
| curlUseHttpsUrl | 12 073 (3,6 %) | 8 401 (2,2 %) | −3 672 | 5 166 (5,1 %) | 5 084 (5,0 %) | −82 |
| apkAddUseNoCache | 11 430 (3,4 %) | 6 278 (1,6 %) | −5 152 | 8 732 (8,7 %) | 4 860 (4,8 %) | −3 872 |
| configureShouldUseBuildFlag | 9 288 (2,7 %) | 9 505 (2,5 %) | 217 | 5 422 (5,4 %) | 5 515 (5,5 %) | 93 |
| wgetUseHttpsUrl | 7 461 (2,2 %) | 7 153 (1,9 %) | −308 | 4 934 (4,9 %) | 4 730 (4,7 %) | −204 |
| tarSomethingRmTheSomething | 5 096 (1,5 %) | 10 841 (2,8 %) | 5 745 | 3 465 (3,5 %) | 7 785 (7,7 %) | 4 320 |
| gpgUseHaPools | 3 338 (1,0 %) | 3 332 (0,9 %) | −6 | 1 155 (1,2 %) | 1 154 (1,1 %) | −1 |
| mkdirUsrSrcThenRemove | 2 666 (0,8 %) | 2 659 (0,7 %) | −7 | 2 458 (2,4 %) | 2 452 (2,4 %) | −6 |
| aptGetInstallUseY | 2 251 (0,7 %) | 5 568 (1,5 %) | 3 317 | 1 741 (1,7 %) | 3 462 (3,4 %) | 1 721 |
| npmCacheCleanUseForce | 1 344 (0,4 %) | 1 354 (0,4 %) | 10 | 1 323 (1,3 %) | 1 333 (1,3 %) | 10 |
| rmRecursiveAfterMktempD | 1 187 (0,3 %) | 397 (0,1 %) | −790 | 296 (0,3 %) | 292 (0,3 %) | −4 |
| sha256sumEchoOneSpaces | 881 (0,3 %) | 2 243 (0,6 %) | 1 362 | 409 (0,4 %) | 1 916 (1,9 %) | 1 507 |
| gemUpdateSystemRmRootGem | 249 (0,1 %) | 246 (0,1 %) | −3 | 231 (0,2 %) | 229 (0,2 %) | −2 |
| gemUpdateNoDocument | 161 (0,0 %) | 158 (0,0 %) | −3 | 145 (0,1 %) | 143 (0,1 %) | −2 |
| gpgVerifyAscRmAsc | 92 (0,0 %) | 85 (0,0 %) | −7 | 90 (0,1 %) | 83 (0,1 %) | −7 |
| yumInstallForceYes | 18 (0,0 %) | 18 (0,0 %) | 0 | 15 (0,0 %) | 15 (0,0 %) | 0 |
| Total | 339 338 | 383 043 | 68 015 | 100 876 | 100 414 | 23 566 |

Table 3 presents the results of the smell repairs. The complete analysis and repair took 1d 14h 57m 44s 576ms which represents an average of 380ms per Dockerfile. The execution includes two AST parsings, two smell analyses, one pretty-print, and one diff computing (between the original Dockerfile and repaired Dockerfile). The first column of Table 3 contains the name of the smell, and the second column contains the number of occurrences of this smell in the two datasets. The third column contains this information after the repair. The fourth and fifth columns contain the same information but instead count the number of Dockerfiles, i.e., a Dockerfile can contain more than one occurrence of a specific smell. Those results show that Parfum is indeed able to repair those Docker smells. However, the rule *aptGetUpdatePreceedsInstall* is by far the less repaired rule. This rule requires deeper refactoring in order to handle the smell properly and not break the build. The supported cases still allow repairing 11 236 cases.

To ensure that Parfum does not break the Docker image build, we built the repaired Dockerfiles. We could not scale the evaluation to the 210 901 Dockerfiles that contains a smell due to the amount of computing it would have required. Indeed, it took us on average 8m 37s to build on Docker image, and it was not possible to increase the parallelization since Dockerhub limits the number of pulls per hour. Based on that information, we concluded that it is not reasonable to build the 210 901 Dockerfiles. Instead, we selected all the Dockerfiles that are located at the root of the repository and that are exactly named `Dockerfile`. We expect that those Dockerfiles are the main Dockerfiles of the repositories and therefore are the most relevant files to repair. We end up with 13 440 Dockerfiles, we build those Dockerfiles, and only 6 507 of them build before the repair. After the repair, 6 138 Dockerfiles build. Therefore, 369 (5,7 %) are failing due to the repair (or build flakiness). It is difficult to estimate the number of builds that are failing due to the build flakiness. However, it is reasonable to believe that the majority is due to Parfum repairs.

Table 4 shows the number of builds that finish with an error per smell type. Note, we consider that all applied repairs have impacted the build status. We observe that the vast majority of the errors are related to the rules *aptGetInstallUseNoRec* and *aptGetInstallThenRemoveAptLists*. Those rules can break builds when a recommended package is removed when it is required or when Parfum removes the cache when it was already empty.

In a few cases, Parfum produces invalid repairs such as Listing 4. In this case, Parfum places `rm gsl.tgz` after the change of directory (`cd gsl-1.16`), the file `gsl.tgz` is, therefore, not found and the build fails. This error could be addressed in Parfum with more advance analysis.

**Answer to RQ2**. Parfum is able to repair the vast majority of the smell (91,1 %) and it only breaks 5,7 % of the builds. It shows that Parfum does not have a 100% success rate however the success rate is high and Parfum can be used to propose changes to developers as we did in RQ4 (see Section 4.6).

**Table 3: The number of occurrences of each smell before and after the repair by Parfum on the two datasets.**

| Dockerfile with Smell | # Docker Smell | | # Dockerfile with Smell | |
|---|---|---|---|---|
| | Before Repair | After Repaired | Before Repair | After Repaired |
| aptGetInstallUseNoRec | 188 801 (23,4 %) | 1 (0,0 %) | 92 290 (43,8 %) | 1 (0,0 %) |
| aptGetInstallThenRemoveAptLists | 161 889 (20,1 %) | 1 940 (2,7 %) | 85 259 (40,4 %) | 477 (0,1 %) |
| curlUseFlagF | 83 498 (10,4 %) | 244 (0,3 %) | 58 303 (27,6 %) | 149 (0,0 %) |
| pipUseNoCacheDir | 78 752 (9,8 %) | 8 (0,0 %) | 43 596 (20,7 %) | 4 (0,0 %) |
| aptGetUpdatePrecedesInstall | 76 643 (9,5 %) | 65 407 (91,6 %) | 36 694 (17,4 %) | 25 655 (7,0 %) |
| npmCacheCleanAfterInstall | 33 870 (4,2 %) | 117 (0,2 %) | 25 699 (12,2 %) | 99 (0,0 %) |
| gpgUseBatchFlag | 32 285 (4,0 %) | 307 (0,4 %) | 11 940 (5,7 %) | 142 (0,0 %) |
| yumInstallRmVarCacheYum | 26 840 (3,3 %) | 180 (0,3 %) | 13 472 (6,4 %) | 92 (0,0 %) |
| tarSomethingRmTheSomething | 24 710 (3,1 %) | 424 (0,6 %) | 17 932 (8,5 %) | 332 (0,1 %) |
| configureShouldUseBuildFlag | 23 675 (2,9 %) | 66 (0,1 %) | 13 426 (6,4 %) | 53 (0,0 %) |
| apkAddUseNoCache | 16 548 (2,1 %) | 13 (0,0 %) | 12 782 (6,1 %) | 9 (0,0 %) |
| wgetUseHttpsUrl | 13 340 (1,7 %) | 0 (0,0 %) | 9 547 (4,5 %) | 0 (0,0 %) |
| curlUseHttpsUrl | 11 045 (1,4 %) | 0 (0,0 %) | 6 992 (3,3 %) | 0 (0,0 %) |
| aptGetInstallUseY | 9 162 (1,1 %) | 5 (0,0 %) | 6 228 (3,0 %) | 4 (0,0 %) |
| yarnCacheCleanAfterInstall | 7 460 (0,9 %) | 2 547 (3,6 %) | 5 320 (2,5 %) | 1 512 (0,4 %) |
| mkdirUsrSrcThenRemove | 5 342 (0,7 %) | 47 (0,1 %) | 4 866 (2,3 %) | 45 (0,0 %) |
| sha256sumEchoOneSpaces | 4 170 (0,5 %) | 2 (0,0 %) | 3 405 (1,6 %) | 2 (0,0 %) |
| gpgUseHaPools | 3 691 (0,5 %) | 0 (0,0 %) | 1 410 (0,7 %) | 0 (0,0 %) |
| npmCacheCleanUseForce | 2 541 (0,3 %) | 8 (0,0 %) | 2 500 (1,2 %) | 8 (0,0 %) |
| rmRecursiveAfterMktempD | 823 (0,1 %) | 112 (0,2 %) | 529 (0,3 %) | 110 (0,0 %) |
| gemUpdateSystemRmRootGem | 527 (0,1 %) | 3 (0,0 %) | 476 (0,2 %) | 3 (0,0 %) |
| gemUpdateNoDocument | 402 (0,0 %) | 0 (0,0 %) | 356 (0,2 %) | 0 (0,0 %) |
| gpgVerifyAscRmAsc | 175 (0,0 %) | 0 (0,0 %) | 162 (0,1 %) | 0 (0,0 %) |
| yumInstallForceYes | 56 (0,0 %) | 0 (0,0 %) | 51 (0,0 %) | 0 (0,0 %) |
| Total | 806 245 | 71 431 | 210 901 | 369 073 |

**Table 4: The number of build errors per smell.**

| Docker Smell | # Build Errors |
|---|---|
| aptGetInstallUseNoRec | 312 (84,6 %) |
| aptGetInstallThenRemoveAptLists | 254 (68,8 %) |
| pipUseNoCacheDir | 115 (31,2 %) |
| curlUseFlagF | 55 (14,9 %) |
| aptGetUpdatePrecedesInstall | 53 (14,4 %) |
| tarSomethingRmTheSomething | 48 (13,0 %) |
| configureShouldUseBuildFlag | 39 (10,6 %) |
| curlUseFlagL | 35 (9,5 %) |
| npmCacheCleanAfterInstall | 32 (8,7 %) |
| wgetUseHttpsUrl | 25 (6,8 %) |
| apkAddUseNoCache | 17 (4,6 %) |
| aptGetInstallUseY | 7 (1,9 %) |
| curlUseHttpsUrl | 6 (1,6 %) |
| mkdirUsrSrcThenRemove | 6 (1,6 %) |
| yumInstallRmVarCacheYum | 4 (1,1 %) |
| yarnCacheCleanAfterInstall | 3 (0,8 %) |
| npmCacheCleanUseForce | 2 (0,5 %) |
| gpgUseBatchFlag | 1 (0,3 %) |
| sha256sumEchoOneSpaces | 1 (0,3 %) |
| gemUpdateSystemRmRootGem | 1 (0,3 %) |
| gemUpdateNoDocument | 1 (0,3 %) |

```
RUN wget -O gsl.tgz ftp://ftp.gnu.org/gsl-1.16.tar
  && tar -zxf gsl.tgz && mkdir gsl
  && cd gsl-1.16 && ./configure --prefix=/app/gsl
  && make && make install
  && rm gsl.tgz # Added line
```

**Listing 4: Example of invalid repair made by Parfum for the repository github.com/olavolav/te-causality.**

## 4.5 RQ3: Impact of Docker Smells on Docker Image Size

In this third research question, we evaluate the impact of Docker smells on the size of the Docker image. To do this, we built 13 440 Docker images before and after the smell repair made by Parfum. Once the images are built, we can compute the uncompressed size of the images before and after the repair and then compare them.

The results of this research question are presented in Table 5. The first column lists the name of the rule, followed by the total savings, average, median, and maximum savings. The three last columns contain information about the Docker images that are accessible on DockerHub. We provide first the total number of image downloads, the average number of downloads per week since the image has been pushed, and finally the estimation of the downloaded amount of data per week. The smells are sorted from the most impactful smell to the least impactful. It's important to note that while we can observe the space savings per Dockerfile, we are not able to determine the exact space savings for each rule. Moreover, not all Dockerfiles are associated with a DockerHub repository. Therefore, the results presented in this table provide a relative comparison between the rules and not an absolute value of Parfum impact.

**Table 5: The image size reduction per rule, note that the saving is computed at the image level where several smells could have been repaired. Additionally, we do not have the number of downloads for each Docker image, we only have this information for images that share the same name on GitHub and DockerHub.**

| Docker Smell | # Smell | Image Size Reduction | | | | # Docker Pull | | Data saved |
|---|---|---|---|---|---|---|---|---|
| | | Total | Average | Median | Maximum | Total | Per Week | per week |
| aptGetInstallUseNoRec | 2 242 | 188,1 GB (6.2%) | 85,9 MB (6.2%) | 17,1 MB (1.8%) | 4,4 GB (87.7%) | 1 853 687 737 | 6 115 092 | 32,35 TB |
| pipUseNoCacheDir | 2 008 | 180,8 GB (7.1%) | 92,2 MB (7.1%) | 14,6 MB (1.6%) | 6,7 GB (88.3%) | 824 217 886 | 3 527 330 | 8,46 TB |
| aptGetInstallThenRemoveAptLists | 2 170 | 163,2 GB (5.7%) | 77 MB (5.7%) | 13,4 MB (1.4%) | 4,4 GB (87.7%) | 796 638 411 | 3 600 656 | 12,2 TB |
| aptGetUpdatePrecedesInstall | 658 | 67,8 GB (7%) | 105,6 MB (7%) | 22,2 MB (2.1%) | 3,1 GB (84.2%) | 348 825 356 | 1 916 933 | 2,41 TB |
| npmCacheCleanAfterInstall | 1 640 | 46,5 GB (3.6%) | 29 MB (3.6%) | 1,3 KB (0%) | 6,7 GB (88.3%) | 629 022 885 | 2 287 838 | 2,86 TB |
| curlUseFlagF | 740 | 29,4 GB (3.1%) | 40,7 MB (3.1%) | 3,2 KB (0%) | 1,8 GB (87.7%) | 1 241 099 452 | 3 549 957 | 8,76 TB |
| tarSomethingRmTheSomething | 184 | 4,6 GB (2.5%) | 25,7 MB (2.5%) | 318 Bytes (0%) | 383 MB (38.7%) | 155 814 317 | 436 029 | 41,45 GB |
| curlUseFlagL | 65 | 4,6 GB (6.1%) | 72,6 MB (6.1%) | 2,4 MB (0.4%) | 1,2 GB (84.2%) | 866 702 590 | 2 152 890 | 264,55 GB |
| yumInstallRmVarCacheYum | 89 | 4,5 GB (4.8%) | 51,2 MB (4.8%) | 177 Bytes (0%) | 801,4 MB (49.6%) | 182 210 | 666 | 10,23 GB |
| gpgUseBatchFlag | 23 | 3,8 GB (7.2%) | 170,5 MB (7.2%) | 72 Bytes (0%) | 2,2 GB (37.3%) | 15 579 | 141 | 356,17 MB |
| configureShouldUseBuildFlag | 143 | 3,7 GB (3.2%) | 26,2 MB (3.2%) | 316 Bytes (0%) | 2 GB (25.7%) | 4 923 750 | 17 371 | 462,6 MB |
| aptGetInstallUseY | 119 | 3,5 GB (2.6%) | 30,2 MB (2.6%) | 3,4 MB (0.4%) | 381,3 MB (60.7%) | 28 440 783 | 98 606 | 88,75 GB |
| apkAddUseNoCache | 887 | 3,5 GB (1.5%) | 4 MB (1.5%) | 628 Bytes (0%) | 369,5 MB (32.7%) | 634 970 465 | 2 021 951 | 656,27 GB |
| mkdirUsrSrcThenRemove | 219 | 2,6 GB (1.1%) | 12,2 MB (1.1%) | 275 Bytes (0%) | 205,4 MB (32.7%) | 69 452 883 | 226 719 | 1,02 TB |
| wgetUseHttpsUrl | 39 | 1,3 GB (2.2%) | 34,6 MB (2.2%) | 649 Bytes (0%) | 582,2 MB (37%) | 3 662 789 | 9 880 | 370,47 MB |
| curlUseHttpsUrl | 29 | 927 MB (1.7%) | 32 MB (1.7%) | 16 Bytes (0%) | 404 MB (84.2%) | 1 028 378 | 3 188 | 21,15 GB |
| gemUpdateSystemRmRootGem | 34 | 877,4 MB (2.8%) | 25,8 MB (2.8%) | 306 Bytes (0%) | 279,1 MB (17.6%) | 2 913 371 | 8 585 | 2,6 GB |
| gemUpdateNoDocument | 31 | 863,5 MB (3%) | 27,9 MB (3%) | 948 Bytes (0%) | 279,1 MB (17.6%) | 58 256 | 262 | 2,6 GB |
| npmCacheCleanUseForce | 3 | 56,4 MB (10.1%) | 18,8 MB (10.1%) | 25 Bytes (0%) | 56,4 MB (18.6%) | N.A. | N.A. | N.A. |
| sha256sumEchoOneSpaces | 14 | 49,1 MB (0.6%) | 3,5 MB (0.6%) | 46 Bytes (0%) | 16,4 MB (11.6%) | 121 674 777 | 421 642 | 2,06 TB |
| rmRecursiveAfterMktempD | 2 | 43 Bytes (0%) | 21,5 Bytes (0%) | N.A. (0%) | 43 Bytes (0%) | 37 128 438 | 310 272 | 3,98 MB |
| gpgUseHaPools | 2 | 6 Bytes (0%) | 3 Bytes (0%) | N.A. (0%) | 6 Bytes (0%) | N.A. | N.A. | N.A. |
| Total | | 275,76 GB | 46,38 MB | 1,51 MB | 6,66 GB | 4 065 282 200 | 13 068 731 | 39,93 TB |

In total, Parfum could save 275,76 GB across 6 138 Dockerfiles. On average, Parfum saved 46,38 MB and on median 1,51 MB. We notice that the most impactful smells are related to packaging managers, developers forget to remove the cache, e.g., *aptGetInstallThenRemoveAptLists*, *pipUseNoCacheDir*, *npmCacheCleanAfterInstall*, *aptGetInstallUseNoRec*. Interestingly, those smells are also among the most frequent smells and the easiest to fix.

When considering the number of times each Docker image is downloaded on Dockerhub (not all Docker images are uploaded to Dockerhub), the amount of data that Parfum can save is significant. Based on the 1 405 repaired Docker images that we found on DockerHub, we estimate a total savings of 39,93 TB of transfer per week. The estimation is based on the total number of downloads since the image has been created on Dockerhub and the size difference between the original Docker image and the repaired Docker image divided by the median image compression ratio (3.2x) reported by Zhao et al. [28]. This highlights the importance of bringing awareness to developers about these practices and reducing the data used for the images. Smaller images should also reduce the build time (we tried to measure this aspect unfortunately the fluctuation due to the network made it complicated to have an accurate measurement) and improve the responsiveness of services that rely on Docker.

Overall, our results show that the impact of Docker smells on the size of the Docker images is significant and that Parfum is effective in repairing these smells and reducing the size of the images.

**Answer to RQ3**. Docker smells have a significant impact on the size of Docker images and using Parfum can lead to a significant reduction of the Docker image without affecting the behavior of the Docker image. On average, the application of Parfum leads to a reduction in image size by 46,38 MB and a total of 1,37 TB of transfer per week (on DockerHub). The most impactful smells identified in this study were related to the use of packaging manager commands, which are also some of the most frequent smells found in the Dockerfiles.

## 4.6 RQ4: Attitudes towards Repairing Docker Smells

In this final research, we investigate developers' attitudes toward the repair of code smells in their projects. To do this, we submitted pull requests containing the output generated by Parfum. We chose not to automatically open pull requests on the projects under study in order to avoid generating unnecessary noise for those projects. Additionally, the process of opening pull requests required manual tasks such as identifying the development branch, determining whether the project accepts external contributions, and ensuring compliance with the project's contribution guidelines.

In this evaluation, we submitted a total of 35 pull requests all of them on different organizations. A list of the opened pull requests can be found in the artifact associated with this repository [6]. The selection criteria to open the pull requests are the following: (1) Dockerfile has at least one smell and less than ten. (2) The repository is active (not achieved, not a fork, has issues, has at least a fork, and has a commit in the last two months on the main branch). (3) The Docker image builds after repair. (4) No more than one pull request per GitHub organization. (5) The image has to be downloaded at least 1 000 from Dockerhub. (6) The size difference needs to be bigger than 1 Mb. Based on those criteria, we obtain a list of 124 candidates and we hand picket 35 repositories that do explicitly welcome external contributions. From those of 35, 20 (57,1 %) were accepted and merged, one of them required manual change and none have been refused. The other 15 (42,9 %) pull requests waiting for an answer from the developers. We selected projects that had an activity in the last two months in order to increase the likelihood of obtaining an answer.

**Table 6: List of the pull requests that receive an answer from the maintainers. The complete list of opened pull requests is available in our repository [6].**

| Project | # Stars | # Image Pull | | Image Size | PR ID | Status | # Smell | Data Saved | |
|---|---|---|---|---|---|---|---|---|---|
| | | Total | Per Week | | | | | Image | per Week |
| AdWerx/pronto-ruby | 19 | 198 057 | 1 125 | 857,35 MB | 171 | Merged | 6 | 38.68 MB (4.51%) | 13,28 GB |
| pelias/openaddresses | 41 | 90 188 | 304 | 577,31 MB | 514 | Merged | 2 | 131.65 MB (22.8%) | 12,2 GB |
| TomWright/mermaid-server | 178 | 2 172 | 15 | 889,97 MB | 122 | Merged | 2 | 28.56 MB (3.21%) | 136,32 MB |
| sqlfluff/sqlfluff | 5576 | 43 313 | 743 | 208,15 MB | 4262 | Merged | 3 | 11.74 MB (5.64%) | 2,66 GB |
| rchakode/realopinsight | 58 | 26 023 | 118 | 809 MB | 30 | Merged | 2 | 39 MB (4.82%) | 1,4 GB |
| vyperlang/vyper | 4422 | 72 697 | 443 | 453,77 MB | 3224 | Merged | 1 | 23.9 MB (5.27%) | 3,23 GB |
| Kruptein/PlanarAlly | 334 | 165 010 | 848 | 342,55 MB | 1142 | Merged | 3 | 31.07 MB (9.07%) | 8,04 GB |
| ShaneIsrael/fireshare | 368 | 10 968 | 340 | 879,21 MB | 166 | Merged | 4 | 158.71 MB (18.05%) | 16,45 GB |
| jcraigk/kudochest | 18 | 2 096 | 28 | 1,52 GB | 187 | Merged | 3 | 302.91 MB (19.42%) | 2,57 GB |
| fzls/djc_helper | 245 | 6 682 | 95 | 489,2 MB | 149 | Merged | 4 | 266.1 MB (54.39%) | 7,68 GB |
| gotzl/accservermanager | 42 | 7 040 | 35 | 1,14 GB | 53 | Merged | 1 | 680.38 MB (58.07%) | 7,32 GB |
| nitrictech/cli | 19 | 1 042 | 34 | 1,47 GB | 438 | Merged | 4 | 113.78 MB (7.55%) | 1,19 GB |
| artsy/hokusai | 87 | 396 984 | 1 442 | 539,42 MB | 323 | Merged | 2 | 10.07 MB (1.87%) | 4,43 GB |
| brndnmtthws/tweet-delete | 77 | 14 727 | 74 | 478,49 MB | 107 | Merged | 2 | 19.94 MB (4.17%) | 460,73 MB |
| bitovi/bitops | 28 | 8 496 | 70 | 168,45 MB | 390 | Merged | 2 | 12.31 MB (7.31%) | 270,08 MB |
| evennia/evennia | 1544 | 37 540 | 121 | 1,25 GB | 3091 | Merged | 5 | 195.49 MB (15.24%) | 7,22 GB |
| sbs20/scanservjs | 436 | 253 233 | 1 846 | 1,04 GB | 527 | Merged | 6 | 419.13 MB (39.45%) | 236,15 GB |
| lncapital/torq | 104 | 2 678 | 62 | 149,25 MB | 307 | Merged | 4 | 17.02 MB (11.4%) | 331,97 MB |
| mitre/saf | 79 | 4 113 | 75 | 603,93 MB | 989 | Merged | 2 | 124.01 MB (20.53%) | 2,83 GB |
| w9jds/firebase-action | 794 | 3 167 517 | 28 147 | 1,36 GB | 176 | Merged | 4 | 124.99 MB (8.96%) | 1,05 TB |
| 35 Opened, 20 (57,1 %) Merged, 0 (0,0 %) Closed, 15 (42,9 %) Pending Pull Requests | | | | | | | 62 | 2,68 GB | 1,37 TB |

Table 6 presents the pull requests that received an answer. The first line contains the repository name, the second contains the number of stars, followed by the total and average weekly downloads on Dockerhub. We then have the original size of the image, the pull request id, the pull request status, the number of repaired smells and the image size reduction, and finally the theoretical average bandwidth saving per week (considering a median compression rate of 3,2 [28]). In total the accepted pull requests resulted in a saving of 2,68 GB which translates to a saving of 1,37 TB per week when we consider the 35 964 average downloads per week.

The pull requests did not elicit significant discussion; most developers simply thanked us for the contributions and proceeded to merge the pull requests. However, some developers were interested to include those changes in other repositories or including the change in the base image of the project. Other developers also thanked us for teaching them something new. Others were curious to know if the repair was automated in some way.

Overall, our results suggest that developers are open and even welcome the repairs of code smells in their projects. This shows that automatizing Docker smell repairs are feasible and that should continue to be explored in the future with the support of additional rules. Additionally, the few discussions that we had with the developers led to a better understanding of what can be done by the developers to improve their Docker image which is particularly rewarding since educating developers about best practices for Dockerfile development was one of the goals of this contribution.

**Answer to RQ4**. We submitted 35 pull requests, 20 have been accepted, none has been rejected. We have been able to achieve a total savings of 2,68 GB and weekly savings of 1,37 TB in bandwidth. This suggests that developers are open to repairing code smells in their projects and that Parfum is able to generate patches that are relevant for developers.

## 5 RELATED WORK

Docker has become a popular tool for developers and organizations to package, deploy, and run applications in a lightweight, portable container. As such, there has been a significant amount of research focused on improving the efficiency, security, and maintainability of Docker-based projects. In this section, we review several relevant studies that are related to this contribution.

A large number of papers studied the Docker ecosystem. We present a selection of them. Ibrahim et al. [15] investigate the number and diversity of images available on Docker Hub for the same system, finding that there is a large number of images to choose from and significant differences between them. Ksontini et al. [16] study the occurrence of refactorings and technical debt in Docker projects, finding that refactorings are common but technical debt is rare. Xu et al. [26] present a study of mining container image repositories for software configuration information, finding that such information is often incomplete or outdated. Lin et al. [18]

study the Docker images hosted on DockerHub. They observe a downward trend of Docker image sizes and smells in Dockerfiles. However, they also observed an upward trend in using obsolete base images. Lui et al. [19] also study DockerHub but focused on the security risks associated with it. Eng et al. [8] did a longitudinal study of the evolution of Dockerfiles, and they confirm that there are slightly fewer smells over time. However, none of those papers study the impact of the smells on the Docker image size.

Other works also focus on improving the security of containers, such as SPEAKER [17], which reduces the number of available systems calls to a given application container by customizing and differentiating its necessary system calls at the booting and the running phases. Confine [9] is a similar technique that uses static analysis to identify the required system calls.

Other contributions aim to improve or fix Dockerfiles. Henkel et al. [14] propose an approach for repairing Dockerfiles that do not build correctly. It uses machine learning to infer repair rules based on build log analysis. Hassan et al. [12] present Rudsea, a technique that adapts Dockerfiles based on the changes in the rest of the project. And Zhang et al. [27] propose a technique that recommends Docker base images to improve efficiency and maintainability. Other tools aim to reduce the size of the Docker images by identifying bloat in the images and removing it. Cimplifier [20] and their framework [21] aims automatically partitions containers into simpler containers based on user-defined constraints. The goals are isolation of each sub-container, communicating as necessary, and only including enough resources to perform their functionality. strip-docker-image [23], minicon [10] and docker-slim [22] are open-source projects that reduce Docker image size by specializing the container to the application.

An important part of the bloat comes from bad practices. Several tools and works focus on identifying those Docker smells. Binnacle [13] is a tool for detecting Docker smells, they compared the presence of those smells between a set of Dockerfiles from GitHub and a set of Dockerfiles written by experts. They observed that there are five times fewer smells in the export Dockerfiles. Wu et al. [24] study the docker smell occurrence in 6 334 projects. They show that smells are very common and there exists co-occurrence between different smells. Xu et al. [25] propose a technique based on static and dynamic analysis to detect temporary files inside Dockerfiles. Nonacademic works focus on detecting Dockerfile smells: Hadolint [11], dockerfilelint [3], docker-bench-security [2], or dockle [4]. However, none of these tools aim to repair the detected smells or aim to reduce the image size by removing Docker smells.

There has been a significant amount of research focused on Docker, including tools for debloating, optimizing, and securing containers, as well as studies of the evolution and management of Dockerfiles and images. Parfum builds upon the related work and introduces Docker smell repair as well as a better understanding of the impact on the Docker image size.

## 6 THREAT TO VALIDITY

One potential threat to validity is the presence of internal bugs in Parfum. A potential threat to validity is the external dependencies of Parfum. Parfum relies on several external libraries for tasks such as parsing and modifying Dockerfiles, and these libraries may have their own bugs or limitations. To mitigate this threat, we tested extensively Parfum and made it open-source.

Another potential threat to validity is the limited scope of Parfum. Parfum only supports a limited number of smells, the supported smells could potentially not be representative of all the smells in Dockerfiles. We mitigate this threat by verifying if the smells are still present in recent Dockerfiles and by opening pull requests to assess the relevance of the smells. Additionally, Parfum only supports parsing and modifying Dockerfiles written in bash and does not support Dockerfiles written in PowerShell or other languages. However, the majority of Dockerfiles are written in bash, therefore, the results are still relevant for the vast majority of users.

An additional potential threat to validity is the diversity of our dataset. While we have collected a large and diverse dataset of Dockerfiles, it is still possible that our results may not generalize to all Dockerfiles or may be biased in some way for example we do not include industry Dockerfiles. This threat to validity also applied to our selection of pull requests. To mitigate this threat, we use an existing dataset (Binnacle one) and created a new one in other to have recent Dockerfiles. Regarding the pull request, we provided the selection criteria and listed the pull requests to be reviewed.

## 7 CONCLUSION

In this paper, we propose Parfum, a tool that automatically detects and repairs Docker smells while producing minimal patches. Parfum is based on a new AST parsing tool, Dinghy, which allows easy modification of Dockerfiles without affecting the file formatting.

We evaluate the effectiveness of Parfum by analyzing and repairing 369 073 Dockerfiles and show that Parfum is able to detect and repair the vast majority of them. This finding is confirmed on our ground trust dataset where we obtain an average precision of 0,96 and recall of 0,95. We also show that Parfum is only breaking 369 (5,7 %) builds and save on average 46,38 MB per Dockerfile.

We concluded our evaluation by proposing the repair to 35 open-source projects. In order to investigate developers' attitudes toward the repair of smells. We found that the developers were positive about the changes and thanked us on several occasions for learning something. 20 pull requests were accepted and merged, one required some manual changes and none were refused.

Overall, our work shows that Parfum is able to detect and automatically repair Dockerfile smells that are frequently present in Dockerfiles. Our approach is well accepted by developers and our tool is easy to use for developers and organizations.

In future work, we plan to expand the capabilities of Parfum to handle more Docker smells, handling more advanced analysis and refactoring. Finally, we would like to integrate Parfum in popular IDE to increase awareness of Docker smells.

## DATA AVAILABILITY

We provide the scripts, dataset, and tool used in this contribution. You can find Dinghy at [5], Parfum at [7], and the evaluation data at [6] on GitHub as well as a functional demo of Parfum at https://durieux.me/docker-parfum.

## REFERENCES

[1] Ryan Chamberlain and Jennifer Schommer. 2014. Using Docker to support reproducible research. *DOI: https://doi. org/10.6084/m9. figshare* 1101910 (2014), 44.

[2] docker-bench security. 2022. docker-bench-security: script that checks Docker deployment best practices. https://github.com/docker/docker-bench-security.
[3] dockerfilelint. 2020. hadolint: An opinionated Dockerfile linter. https://github.com/replicatedhq/dockerfilelint.
[4] dockle. 2020. dockle: Container Image Linter for Security. https://github.com/goodwithtech/dockle.
[5] Thomas Durieux. 2023. Open-science repository for Dinghy. Online repository: https://github.com/tdurieux/Dinghy.
[6] Thomas Durieux. 2023. Open-science repository for the experiments of Parfum. Online repository: https://github.com/tdurieux/docker-parfum-experiment.
[7] Thomas Durieux. 2023. Open-science repository for Parfum. Online repository: https://github.com/tdurieux/docker-parfum.
[8] Kalvin Eng and Abram Hindle. 2021. Revisiting Dockerfiles in Open Source Software Over Time. In *2021 IEEE/ACM 18th International Conference on Mining Software Repositories (MSR)*. IEEE, New York, NY, USA, 449–459.
[9] Seyedhamed Ghavamnia, Tapti Palit, Azzedine Benameur, and Michalis Polychronakis. 2020. Confine: Automated System Call Policy Generation for Container Attack Surface Reduction. In *23rd International Symposium on Research in Attacks, Intrusions and Defenses (RAID 2020)*. USENIX Association, San Sebastian, 443–458. https://www.usenix.org/conference/raid2020/presentation/ghavanmnia
[10] grycap. 2020. Minimization of the filesystem for containers. https://github.com/grycap/minicon.
[11] hadolint. 2022. hadolint: Dockerfile linter validate inline bash written in haskell. https://github.com/hadolint/hadolint.
[12] Foyzul Hassan, Rodney Rodriguez, and Xiaoyin Wang. 2018. RUDSEA: Recommending Updates of Dockerfiles via Software Environment Analysis. In *Proceedings of the 33rd ACM/IEEE International Conference on Automated Software Engineering* (Montpellier, France) *(ASE 2018)*. Association for Computing Machinery, New York, NY, USA, 796–801. https://doi.org/10.1145/3238147.3240470
[13] Jordan Henkel, Christian Bird, Shuvendu K. Lahiri, and Thomas Reps. 2020. Learning from, Understanding, and Supporting DevOps Artifacts for Docker. In *Proceedings of the ACM/IEEE 42nd International Conference on Software Engineering* (Seoul, South Korea) *(ICSE '20)*. Association for Computing Machinery, New York, NY, USA, 38–49. https://doi.org/10.1145/3377811.3380406
[14] Jordan Henkel, Denini Silva, Leopoldo Teixeira, Marcelo d'Amorim, and Thomas Reps. 2021. Shipwright: A Human-in-the-Loop System for Dockerfile Repair. In *2021 IEEE/ACM 43rd International Conference on Software Engineering (ICSE)*. IEEE, New York, NY, USA, 1148–1160.
[15] Md Hasan Ibrahim, Mohammed Sayagh, and Ahmed E. Hassan. 2020. Too many images on DockerHub! How different are images for the same system? *Empir. Softw. Eng.* 25, 5 (2020), 4250–4281.
[16] Emna Ksontini, Marouane Kessentini, Thiago do N Ferreira, and Foyzul Hassan. 2021. Refactorings and Technical Debt in Docker Projects: An Empirical Study. In *2021 36th IEEE/ACM International Conference on Automated Software Engineering (ASE)*. IEEE, New York, NY, USA, 781–791.
[17] Lingguang Lei, Jianhua Sun, Kun Sun, Chris Shenefiel, Rui Ma, Yuewu Wang, and Qi Li. 2017. SPEAKER: Split-phase execution of application containers. In *International Conference on Detection of Intrusions and Malware, and Vulnerability Assessment*. Springer, New York, NY, USA, 230–251.
[18] Changyuan Lin, Sarah Nadi, and Hamzeh Khazaei. 2020. A large-scale data set and an empirical study of docker images hosted on docker hub. In *2020 IEEE International Conference on Software Maintenance and Evolution (ICSME)*. IEEE, New York, NY, USA, 371–381.
[19] Peiyu Liu, Shouling Ji, Lirong Fu, Kangjie Lu, Xuhong Zhang, Wei-Han Lee, Tao Lu, Wenzhi Chen, and Raheem Beyah. 2020. Understanding the Security Risks of Docker Hub. In *Computer Security – ESORICS 2020*, Liqun Chen, Ninghui Li, Kaitai Liang, and Steve Schneider (Eds.). Springer International Publishing, Cham, 257–276.
[20] Vaibhav Rastogi, Drew Davidson, Lorenzo De Carli, Somesh Jha, and Patrick McDaniel. 2017. Cimplifier: automatically debloating containers. In *Proceedings of the 2017 11th Joint Meeting on Foundations of Software Engineering*. Association for Computing Machinery, New York, NY, USA, 476–486.
[21] Vaibhav Rastogi, Chaitra Niddodi, Sibin Mohan, and Somesh Jha. 2017. New Directions for Container Debloating. In *Proceedings of the 2017 Workshop on Forming an Ecosystem Around Software Transformation* (Dallas, Texas, USA) *(FEAST '17)*. Association for Computing Machinery, New York, NY, USA, 51–56. https://doi.org/10.1145/3141235.3141241
[22] SlimToolkit. 2023. Inspect, Optimize and Debug Your Containers. https://github.com/slimtoolkit/slim.
[23] Mark van Holsteijn. 2018. Utility to strip Docker images to their bare minimum size. https://github.com/mvanholsteijn/strip-docker-image.
[24] Yiwen Wu, Yang Zhang, Tao Wang, and Huaimin Wang. 2020. Characterizing the occurrence of dockerfile smells in open-source software: An empirical study. *IEEE Access* 8 (2020), 34127–34139.
[25] Jiwei Xu, Yuewen Wu, Zhigang Lu, and Tao Wang. 2019. Dockerfile tf smell detection based on dynamic and static analysis methods. In *2019 IEEE 43rd Annual Computer Software and Applications Conference (COMPSAC)*, Vol. 1. IEEE, New York, NY, USA, 185–190.
[26] Tianyin Xu and Darko Marinov. 2018. Mining container image repositories for software configuration and beyond. In *Proceedings of the 40th International Conference on Software Engineering: New Ideas and Emerging Results*. Association for Computing Machinery, New York, NY, USA, 49–52.
[27] Yinyuan Zhang, Yang Zhang, Xinjun Mao, Yiwen Wu, Bo Lin, and Shangwen Wang. 2022. Recommending Base Image for Docker Containers based on Deep Configuration Comprehension. In *2022 IEEE International Conference on Software Analysis, Evolution and Reengineering (SANER)*. IEEE, New York, NY, USA, 449–453. https://doi.org/10.1109/SANER53432.2022.00060
[28] Nannan Zhao, Vasily Tarasov, Hadeel Albahar, Ali Anwar, Lukas Rupprecht, Dimitrios Skourtis, Arnab K Paul, Keren Chen, and Ali R Butt. 2020. Large-scale analysis of docker images and performance implications for container storage systems. *IEEE Transactions on Parallel and Distributed Systems* 32, 4 (2020), 918–930.